\documentclass[onecolumn,showkeys,preprintnumbers,aps,a4paper,amssymb,prd,superscriptaddress,nofootinbib]{revtex4-2}
\usepackage{comment}
\usepackage{graphicx}
\usepackage{epsf}
\usepackage{bm}
\usepackage{amsmath}
\usepackage{amsfonts}
\usepackage{amssymb}
\usepackage{epstopdf}
\usepackage{color}
\usepackage[dvipsnames]{xcolor}
\usepackage{verbatim}
\usepackage{multirow}
\usepackage{soul}
\usepackage{physics}
\usepackage{bm}

\usepackage[width=0.00cm, height=0.00cm, left=1.50cm, right=1.50cm, top=2.00cm, bottom=2.00cm]{geometry}
\usepackage{microtype}
\usepackage{lmodern}

\usepackage[colorlinks = true,
            linkcolor = teal,
            urlcolor  = teal,
            citecolor = teal,
            anchorcolor = blue]{hyperref}
\usepackage[capitalize]{cleveref}
\usepackage[normalem]{ulem}
\usepackage{enumitem}
\usepackage{booktabs}

\usepackage{lipsum}

\makeatletter\let\expandableinput\@@input\makeatother

\begin{document}

\begin{center}
		\vspace{0.4cm} {\large{\bf Cosmological Implications of a New Creation Field in Hoyle–Narlikar Gravity with Bulk Viscous Fluid}} \\
		\vspace{0.4cm}
		\normalsize{ Archana Dixit$^1$, Manish Yadav$^2$, Anirudh Pradhan$^3$, M. S. Barak$^4$ }\\
		\vspace{5mm}
			\normalsize{$^{1}$ Department of Mathematics, Gurugram University, Gurugram, Haryana 122003, India}\\ 
			\vspace{2mm}
		\normalsize{$^{2,4}$ Department of Mathematics, Indira Gandhi University, Meerpur, Haryana 122502, India}\\
		\vspace{2mm}
        \normalsize{$^{3 }$ Centre for Cosmology, Astrophysics and Space Science (CCASS), GLA University, Mathura, Uttar Pradesh, India}\\ 
		\vspace{2mm}
		$^1$Email address: archana.ibs.maths@gmail.com\\
			$^2$Email address: manish.math.rs@igu.ac.in\\
		    $^3$Email address: pradhan.anirudh@gmail.com\\
             $^4$Email address: ms$_{-}$barak@igu.ac.in\\
\end{center}

 
\maketitle
\section{Abstract}

 	In this study, we present a comprehensive investigation of the Narlikar gravity model with bulk viscous fluid by the new foam of creation field $C(t) = t + \int \alpha (1 - a)\,dt + c_1$, based on the  Hoyle–Narlikar’s creation-field theory, using a joint analysis of Observational Hubble Data (OHD) and the Pantheon supernova (PP) compilation. Our analysis reveals that the creation field coupling constant $(f)$  is always positive within the Narlikar gravity model from OHD+PP data sets. The best-fit estimates yield $ H_0 = 71.2 \pm 2.1 \,\text{km s}^{-1}\,\text{Mpc}^{-1}$ and $\xi_0 = 0.23$, quoted at the $1\sigma$ level. The Narlikar gravity model predicts a transition redshift of $z_t \approx 0.63$ marking the onset of late-time cosmic acceleration, with the corresponding age of the Universe estimated as $13.50\pm1.80 \ \  Gyr$. Interestingly, the inferred higher value of $H_0$, relative to SH0ES determinations, suggests a possible alleviation of the $\sim 4.1\sigma$ Hubble tension. Furthermore, we assess the stability of the model and demonstrate that the late-time acceleration can be consistently explained through the energy conditions. This model retains dynamical flexibility while ensuring analytical tractability and provides a promising framework to investigate the cosmological implications of Hoyle–Narlikar gravity, particularly regarding late-time acceleration and the evolution of dark energy. \\
 	
 		\smallskip 
 	{\it Keywords}: New creation field; Hoyle-Narlikar gravity; Bulk viscosity; Energy conditions, Observational constraints\\
 	\smallskip
 	PACS: 98.80.-k, 98.80.Jk

\section{Introduction}

 The modern framework of cosmology rests on the Cosmological Principle, which assumes large-scale homogeneity and isotropy, leading to the Friedmann–Robertson–Walker (FRW) models that form the Standard Model of Cosmology. While this Big Bang–based framework successfully explains key observations such as the Hubble expansion, the Cosmic Microwave Background (CMB), and other datasets, it faces unresolved issues, including the initial singularity, flatness, and horizon problems. These limitations have motivated the exploration of alternative models, such as the Steady State theory proposed by Bondi and Gold \cite{ref1}, which postulates an eternally expanding universe with continuous matter creation at a very low rate. Although this model circumvents the singularity problem, it remains incomplete due to the lack of a satisfactory physical mechanism for continuous creation.\\

 In 1964, a novel creation-field theory was introduced by Hoyle and Narlikar\cite{ref2,ref3,ref4} as an alternative to the standard cosmology model theory based on the Big Bang. This approach modifies the right-hand side of Einstein’s field equations by incorporating a negative-energy scalar field, which acts as a source of continuous matter creation in the universe. Due to this, the expansion of the universe is eternal, but preserving a constant matter density. Further, Narlikar later proposed \cite{ref5} that negative-energy creation fields serve as the fundamental mechanism for matter generation. Moreover, these fields address two of the most persistent challenges in Big Bang cosmology, particularly the horizon and flatness problems (by allowing causal connection across vast regions and driving the universe toward near-flatness). Extending this idea, Hawking \cite{ref6} analyzed the implications of negative mass with creation field under the framework of the Hoyle–Narlikar gravity model. In this context, Narlikar and Padmanabhan \cite{ref7} presented a notable solution to Einstein’s field equations in which radiation is governed by a massless, chargeless scalar field C carrying negative energy. In a subsequent development, Hoyle \cite{ref8} modified the general relativistic field equations to incorporate a continuous matter creation process. This theoretical creation field  approach allows for a stable universe that undergoes expansion without relying on a vacuum foam of dark energy (cosmological constant).\\

McIntosh \cite{ref9} extended the Hoyle–Narlikar theory, which demonstrated its consistency with Mach’s principle, due to this, a more coherent explanation of how local inertial dynamics emerge from the large-scale structure of the universe. Davies \cite{ref10} explored this theory's capability to connect elementary particle physics with cosmological scales, which underscores its significance in unifying fundamental and large-scale phenomena. The conformal invariance transformations of the Hoyle–Narlikar theory was systematically examined their theory,  leading to significant insights into the geometric structure and interpretation of their gravitational model \cite{ref11}. According to the Hoyle–Narlikar theory \cite{ref12} has been extensively applied to a variety of cosmological problems, including static models, and the study of spacetime’s fundamental nature and alternative descriptions of cosmic origin beyond the Big Bang. This theory allows for quantitative restrictions on cosmological parameters, including the Hubble constant and the cosmological constant. As development by Thorne \cite{ref13} explored the framework by examining the role of primordial magnetic fields in cosmological evolution. In parallel, Ref. \cite{ref14} explores the theoretical consequences of dark energy and its contribution to cosmic acceleration in an expanding cosmological background. Cosmological models incorporating a variable cosmological constant and a barotropic fluid within the Hoyle–Narlikar framework have been explored in recent studies \cite{ref15,ref16,ref17} to account for the universe's late-time acceleration. In a related development, Chatterjee and Banerjee \cite{ref18}, within the Hoyle-Narlikar theory, introduced higher-dimensional generalizations of cosmological models. In the framework of quasi-steady state cosmology, Narlikar et al. \cite{ref19} analyzed the gravitational wave background. Furthermore, Hoyle and Narlikar \cite{ref20,ref21,ref22} investigated the role of Mach's principle in matter creation and the implications of baryon non-conservation on cosmology.\\

On the other hand, the inclusion of shear and bulk viscosity within viscous fluid models has been widely investigated and is believed to play a significant role in the cosmic evolution of the universe. The first studies of relativistic viscous fluids, as described in \cite{ref22a,ref22b}, derived parabolic-type equations under the assumption of first-order deviations from equilibrium. However, the results of these equations predicted the infinitely fast propagation of viscous and thermal effects, thereby conflicting with the principle of particle causality. Later, in 1995 \cite{ref22c}, the author presented a second-order deviation from equilibrium and discussed the early as well as late times of the expansion history based on this deviation. In general, viscous fluid processes are characterized by the bulk viscosity parameter denoted as $\xi$, while the contribution of shear viscosity $\eta$ is typically neglected \cite{ref22d}. The bulk viscosity governs the dissipative process in a viscous fluid and modifies the pressure via $p_{\text{eff}} = p-3\xi H$, where $H$ is the Hubble parameter, $P$ is the isotropic pressure, and $\xi$ is the bulk viscous coefficient. The bulk viscosity coefficient must satisfy $\xi > 0$, as explained by the second law of thermodynamics \cite{ref22e,ref2221,refe2} which indicates that entropy production remains positive.  The contribution of bulk viscosity fluid to the universe's accelerated expansion at late times was addressed in \cite{ref22f,ref22ff,ref22h,ref22i}. However, one major drawback of viscous fluid models in an expanding universe is the lack of a well-established origin. The bulk viscosity arises due to the breakdown of local thermodynamic equilibrium, and it acts as an effective pressure tending to return the system to its equilibrium state. This pressure is generated when the cosmological fluid expands or contracts too rapidly, thereby departing from local thermodynamic equilibrium, and vanishes once equilibrium is restored.\\

In the more recent work, D.C. Murya has presented several studies on Hoyle–Narlikar gravity with a novel creation field. In the first study, the author \cite{ref23} examined the role of dark energy and cosmic acceleration with a simple creation field $C(t) \propto t$  from the latest observation data set, such as the cosmic chronometer and Pantheon data. In a subsequent paper, the analysis was extended to a more complicated form of the creation field, $C(t) = t + \int k a^{n}\, da + c_{1}$ \cite{ref24}, where $k$ and $n$ are free parameters of the model, employing the same observational data. After that, three particular cases of this complicated creation field were examined—namely $n= \frac{1}{2}$, $n=1$, and $n=2$ with a dust-dominated universe ($p=0$)— to explore distinct evolutionary behaviors of the universe \cite{ref25}. Motivated by these results, we propose a newly foam creation field, $C(t) = t + \int \alpha(1-a)\, dt + c_{1} $, along with a bulk viscous fluid characterized by $p_{\text{eff}} = p - 3\xi H$, which provides a new mechanism for matter creation in an expanding universe with bulk viscous fluid. This formulation retains dynamical flexibility while ensuring analytical tractability and provides a promising framework to investigate the cosmological implications of Hoyle–Narlikar gravity, particularly regarding late-time acceleration and the evolution of dark energy.\\

The present manuscript is organized into several key sections. Section II provides a comprehensive review of the literature on Hoyle–Narlikar gravity with bulk viscous fluids and introduces a novel form of the creation field. Section III  presents fundamental concepts of the Hoyle–Narlikar theory of gravity and the corresponding field equations in this modified framework. In section  IV, we build up our derived cosmological model and provide a cosmological solution. In the next section V,  we introduce the latest OHD and PP data sets and the methodology employed in our analysis. Section VI presents the analysis results along with a discussion of the key findings. Finally, Section VII provides the concluding remarks and a final summary.

\section{Theory of Holy-Narlikar Gravity with Bulk Viscous Fluid}
\label{sec:datasets}
In the present study, we consider the Hoyle and Narlikar gravity with a bulk viscous fluid. Within this framework, the modified Einstein field equations is given by

\begin{equation}\label{eq1}
R_{\mu\nu} - \frac{1}{2} R g_{\mu\nu} = -8\pi \left[ T_{\mu\nu}  -f\left(  C_{\mu} C_{\nu} - \frac{1}{2} C^{\alpha} C_{\alpha} g_{\mu\nu} \right) \right],
\end{equation}

where \( R \), \( R_{\mu\nu} \), and  \( g_{\mu\nu} \), the Ricci scalar, Ricci curvature tensor, and metric tensor of spacetime geometry, respectively with  $u^{\mu}$ is the four-velocity of the fluid satisfying $u^{\mu} u_{\mu} = 1$, and $C$ is the scalar creation field with  $C_{\mu}=\frac{dC}{dx^{\mu}}$. The parameter $f > 0$ denotes the coupling constant between matter and the creation field. Let us assume that the Universe contains both a normal matter source and a bulk viscous fluid. Then, the energy–momentum tensor $T_{\mu\nu}$ corresponding to the bulk viscous fluid is given by\
\begin{equation}\label{eq2}
	T_{\mu\nu}=(\rho+p_\text{eff})u_{\mu}u_{\nu}-p_\text{eff}\,g_{\mu\nu}
\end{equation}

Here $\rho$ and $p_\text{eff}$ denote the total energy density of the ﬂuid and the effective pressure of the bulk viscous ﬂuid, respectively. The effective pressure of the universe is deﬁned as

\begin{equation}\label{eq3}
	p_\text{eff} = p - 3\xi H
\end{equation} 
where $p$ stands for the normal (equilibrium) pressure of the fluid, and $\xi$ represents the bulk viscous coefficient generated in the fluid due to deviations from local thermal equilibrium. In general, $\xi$ may be considered as a function of the Hubble parameter and its derivatives.\\

 Under this framework, we assume the spacetime geometry of the universe with flat curvature is described by the FRW metric, read as
\begin{equation}\label{eq4}
ds^2 = -c^2 dt^2 + a(t)^2 \left[ {dr^2} + r^2 \left( d\theta^2 + \sin^2\theta \, d\phi^2 \right) \right]
\end{equation}

Where $a(t)$ is the cosmic scale factor change with cosmic time. We derive first and second Friedmann equations from Eq.(\ref{eq1}), Eq.(\ref{eq2}),  Eq.(\ref{eq3}), and Eq.(\ref{eq4}) with bulk viscous fluid coefficient:
\begin{equation}\label{eq5}
	3\mathcal H^2 + \frac{1}{2}f\dot{C}^{2}= \rho
\end{equation}
\begin{equation}\label{eq6}
	2\dot{\mathcal H} + 3\mathcal H^2  -\frac{1}{2}f\dot{C}^{2} =-p + 3\xi H
\end{equation}
 The bulk viscosity coeﬃcient $\xi$ is generally associated with the matter content of the universe as well as the Hubble parameter and its derivative. We assume that $\xi$ depends only on the Hubble parameter, i.e., $\xi = \xi(H)$. In this work, we adopt the following specific functional form of $\xi$:

 $$ \xi = \xi_0 H$$

Where $\xi_0$ is an arbitrary constant. We derive the continuity equation from Eq.(\ref{eq5}) and Eq.(\ref{eq6}) with speciﬁc form  of bulk viscous fluid as

\begin{equation}\label{eq7}
	\dot{\rho}+3H(1+w-\xi_0)\rho=\frac{3}{2}Hf\dot{C}^{2}(2-\xi_0) + f\dot{C}\ddot{C}
\end{equation}
The source term $ 3Hf\dot{C}^{2} + f\dot{C}\ddot{C}$ represents the energy generation or destruction within the system. When $3Hf\dot{C}^{2} + f\dot{C}\ddot{C}= 0 $ the continuity equation Eq.(\ref{eq7}) reduces to the standard continuity equation of General Relativity. However, If $3Hf\dot{C}^{2} + f\dot{C}\ddot{C}\ne0$, then an energy transfer process or particle formation occurs in the system.

\section{Cosmological Solutions}

We investigate the observable behavior of the universe within a generalized Hoyle–Narlikar creation-field theory. Within this theoretical framework, we introduce the concept of the first derivative of the newly defined creation field $C(t)$ with respect to cosmic time, read as
\begin{equation}\label{eq8}
	\dot{C(t)} = \frac{dC(t)}{dt}=1+\alpha(1-a)
\end{equation}
where 
$\alpha>0$ is a free parameter and 
a(t) denotes the scale factor of the universe.

By combining Eqs. (\ref{eq7}), (\ref{eq8}) and using the relation ($\dot{\rho} = \frac{d\rho}{da} \cdot \dot{a}$), we obtain the following first-order linear differential equation for the energy density w.r.t scale factor :
\begin{equation}\label{eq9}
a\frac{d\rho}{da} + {3(1 + \omega -\xi_0)} \rho = f \left[ \frac{3}{2}(2-\xi_0)[1 + \alpha(1 - a)]^2  -a\alpha[1 + \alpha(1 - a)] \right]
\end{equation}
 To solve the linear differential equation w.r.t the scale factor($a$), read as,

\begin{equation}\label{eq10}
\rho(a) = \rho_0\,a^{-3(1+w-\xi_{0})} + \frac{f\left(\tfrac{3}{2}(2-\xi_{0})\right)(1+\alpha)^{2}}{3(1+w-\xi_{0})} - \frac{f\,\alpha(1+\alpha)\left(2\cdot\tfrac{3}{2}(2-\xi_{0})+1\right)}{3(1+w-\xi_{0})+1}\,a + \frac{f\,\alpha^{2}\left(\tfrac{3}{2}(2-\xi_{0})+1\right)}{3(1+w-\xi_{0})+2}\,a^{2}\\
\end{equation}

where $\rho_{0}$ and $a_{0}$ are the present value of density and scale factor of the universe.\\
Substituting Eq.(\ref{eq10}) and Eq.(\ref{eq8})  into Eq.(\ref{eq5}), we express the Hubble parameter in terms of redshift $z$  with  bulk viscous fluid coefficient $\xi_0$ as:
\begin{equation}\label{eq11}
H(z) = H_0 \sqrt{\rho_0\frac{(1 + z)^{3(1+w-\xi_0)}}{3H_0^2} + \frac{ f\alpha^2(1-w)}{2H_0^2{(5+3w-3\xi_0)(1 + z)^2}} + \frac{f\alpha(1 + \alpha) (w-1)}{H_0^2(4+3w -3\xi_0)(1+z)}+ \frac{f(1 + \alpha)^2(1-w)}{6H_0^2(1+w -\xi_0)}  }
\end{equation}

Consequently, the final expression is reduced to the form:
\begin{equation}\label{eq12}
	H(z) = H_0 \sqrt{\Omega_{m0}(1 + z)^{3(1+w-\xi_0)} + \frac{\Omega_{x}}{ (1 + z)^2} +  \frac{\Omega_{y}}{ (1 + z)} +\Omega_z}
\end{equation}
where the density parameters are defined as follows: $\Omega_{m0}=\frac{\rho_{0}}{3H_{0}^{2}}$, $\Omega_{x}=\frac{f \alpha^2}{2H_0^2}\frac{(1 - w)}{(5 + 3w - 3\xi_0 )}$, $\Omega_{y}=\frac{f \alpha (1 + \alpha)}{H_0^2} \frac{(w - 1)}{4+3w-3\xi_0}$, and $\Omega_{z}=\frac{f(1 + \alpha)^2}{6H_0^2}  \frac{1 - w}{(1+w-\xi_0)} $ and they satisfy the closure relation: $\Omega_{m0}+\Omega_{x}+\Omega_{y}+\Omega_{z}=1$.\\

Now, we obtain the effective energy density ($\rho_{eff}$) and effective pressure ($p_{eff}$) as
\begin{equation}\label{eq13}
\rho(z) = 3H_0^2 \left[ \Omega_{m0}(1 + z)^{3(1+ \omega -\xi_0)} + \frac{\Omega_x}{(1 + z)^2} + \frac{\Omega_y}{(1 + z)} + \Omega_z \right] + \frac{1}{2}f \left(1 + \frac{\alpha z}{1 + z} \right)^2
\end{equation}

and
		
\begin{equation}\label{eq14}
p(z) = H_0^2 \left[ 3w\Omega_{m0}(1+z)^{3(1+w-\xi_0)} + (3\xi_0 -5)\frac{\Omega_x}{(1+z)^2} +(3\xi_0 -4) \frac{\Omega_y}{(1+z)} +3(\xi_0-1)\Omega_z \right] + \frac{1}{2}f\left(1 + \alpha \cdot \frac{z}{1+z} \right)^2
\end{equation}
respectively.\\

\section{DATASETS AND METHODOLOGY}

\textbf{Observational Hubble Data }: The Observational Hubble data (OHD)  method relies on estimating the differential ages of the oldest passively evolving galaxies at closely separated redshifts. This approach provides  a model-independent determination of the expansion rate of the Universe by relating the Hubble parameter to the redshift and cosmic time via the relation
$H(z) = -\frac{1}{1+z} \frac{dz}{dt},$ as originally proposed in foundational studies~\cite{ref26}. In this work, we use 33 non-correlated measurements of the Hubble parameter $H(z)$, covering the redshift interval $0.07 \leq z \leq 1.965$ \cite{ref27,ref28,ref29,ref30,ref31,ref32,ref33,ref34,ref34a}, obtained through the various Hubble surveys.\\

\begin{figure}[hbt!]
    \centering
    \includegraphics[width=0.9\linewidth]{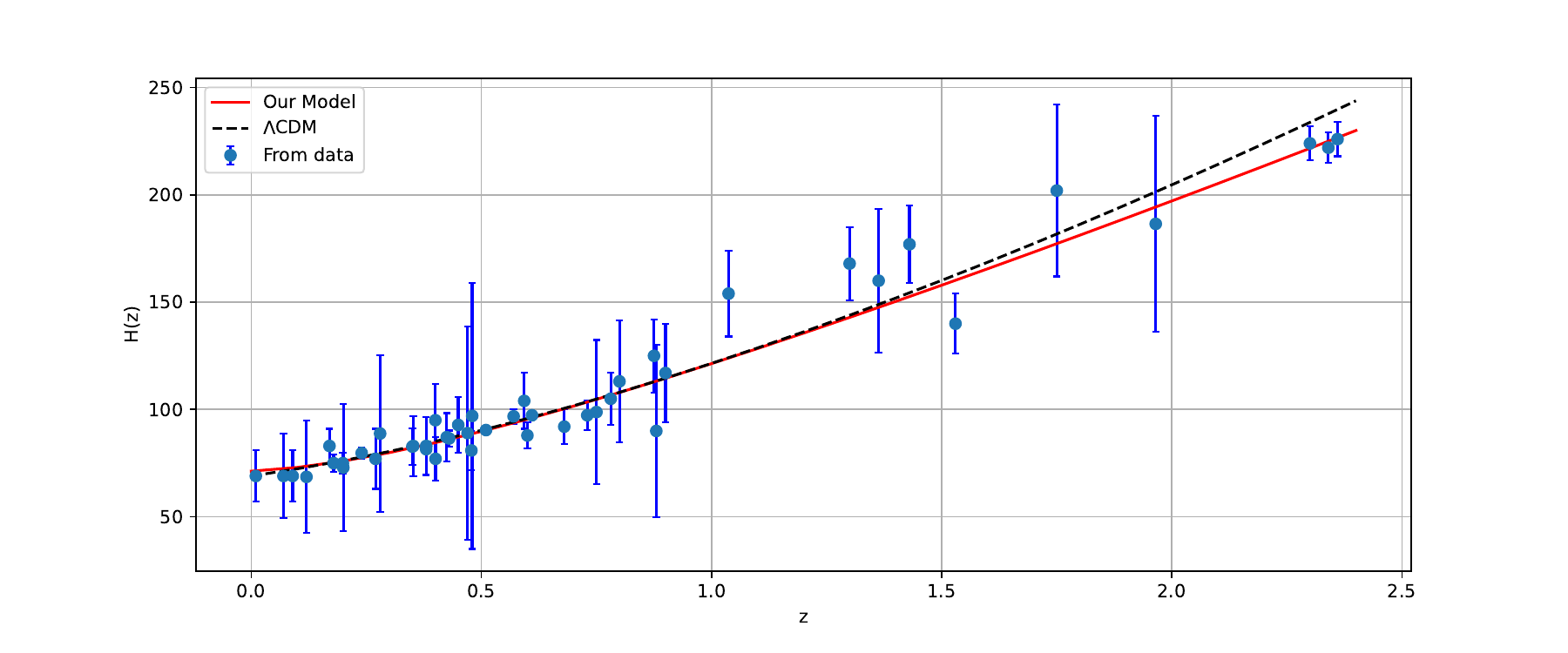}
    \caption{The Hubble parameter with blue colour $1\sigma$ error bar for the combined OHD+PP datasets as a function of redshift z.}
    \label{fig1}
\end{figure}

We will define the chi-squared function, signified by $\chi^2_{\rm OHD}$, for these measurements as follows:

$$\chi^2_{\text{OHD}} = \sum_{i=1}^{33} \frac{\left[ d^{obs}(z_i) - d^{th}(z_i) \right]^2}{\sigma^2_{d^{obs}(z_i)}},$$

Where $ d^{obs}(z_i)$ and $ d^{th}(z_i)$ denotes the observed and model-predicted values of the Hubble parameter at redshift $z_i$, respectively, and $\sigma_{{d^{obs}(z_i)}}$ represents the associated observational uncertainty, as provided in Ref. \\
In our analysis, we employ these measurements to constrain the parameters of the proposed cosmological scenario. The theoretical predictions of the Hubble value from our model-dependent approach show good agreement with the model-independent OHD as well as $\Lambda$CDM model across the redshift range, as illustrated in Fig.\ref{fig1}.\\

\begin{figure}[hbt!]
    \centering
    \includegraphics[width=0.9\linewidth]{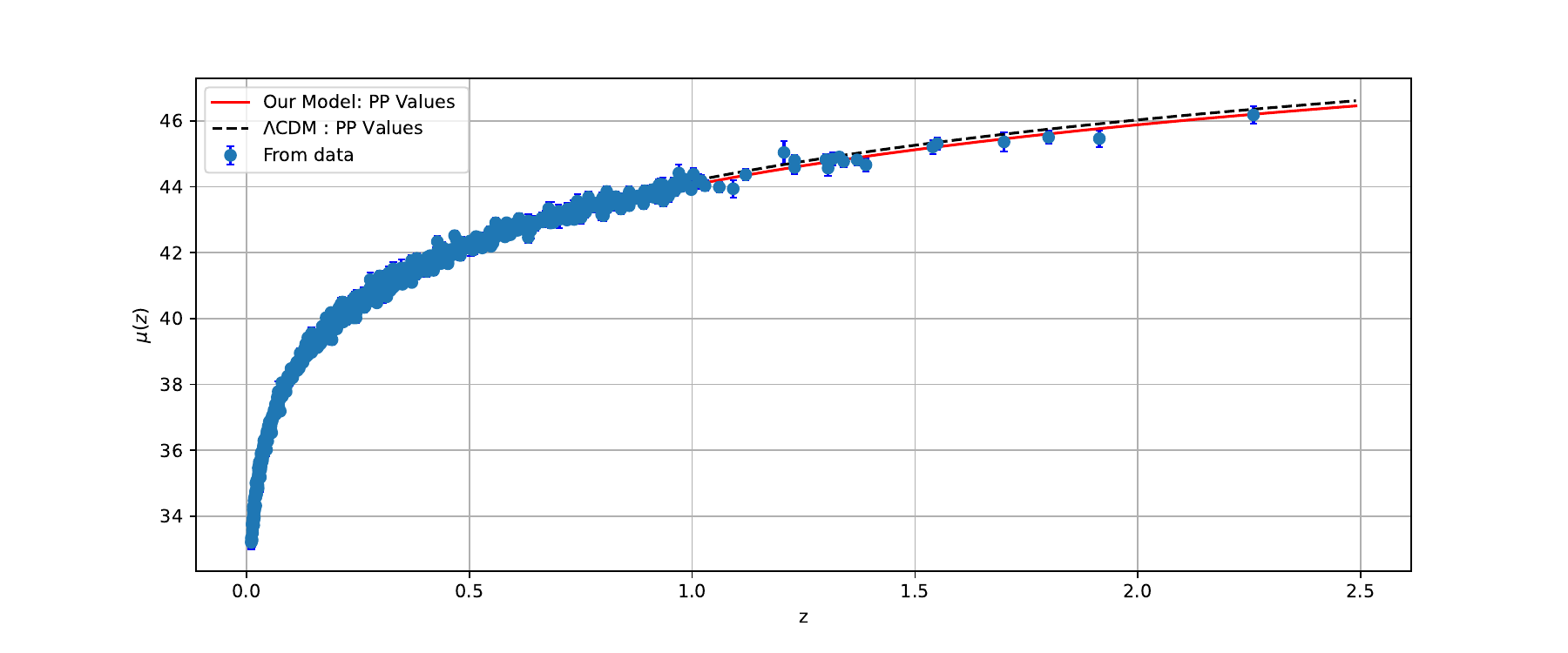}
    \caption{The distance modulus for the combined OHD+PP datasets as a function of redshift z.}
    \label{fig2}
\end{figure}

\textbf{Pantheon}: Type Ia supernovae are widely recognized as powerful standard candles in modern astrophysics cosmology, enabling the reconstruction of cosmic expansion and offering stringent constraints on cosmological parameters of the derived model. In this study, we utilized  Pantheon (PP) compilation sample of 1048 Type Ia supernova data points from references \cite{ref35}, spanning a redshift range of \( z \in [0.001, 2.26]\). These supernovae serve as precise probes for measuring distance moduli, thereby constraining the uncalibrated luminosity distance scaled by the Hubble constant, $H_0 d_L(z)$. Moreover, the distance modulus \( \mu_{\text{th}}(z) \), describing apparent magnitude \( m \) and absolute magnitude \( M_b \), is read as
\begin{equation}\label{eq15}
\mu(z) = -M_b + m = \mu_0 + 5 \log D_L(z), 
\end{equation}
where \( D_L(z) \) and \( \mu_0 \) denote the luminosity distance and the nuisance parameter respectively. Since the absolute magnitude \( M_b \) has a strong correlation with \( H_0 \), it is considered a free parameter.\\

Now, equation \eqref{eq15} leads to
\begin{equation}\label{eq16}
\mu_0 = 5 \log \left( \frac{H_0^{-1}}{1\,\text{Mpc}} \right) + 25,
\end{equation}

Thus, the luminosity distance \( D_L \) is defined as

\begin{equation}\label{eq17}
D_L = (1 + z) \int_0^z \frac{H_0}{H(z')} \, dz'.
\end{equation}
And the corresponding $\chi^2_{\text{PP}}$ can be derived as follows

$$\chi^2_{\text{PP}} = \sum_{i,j}^{1048} \Delta\mu_i \left(C_{\text{stat + syst}}^{-1}\right)_{ij} \Delta\mu_j.$$

Here, $\Delta \mu_i = \mu^{\text{th}}_i - \mu^{\text{obs}}_i$ represents the deviation between the theoretical and observed distance modulus values. $C_{\text{stat + syst}}^{-1}$ is the inverse of the covariance matrix corresponding to the Pantheon dataset, which accounts for statistical and systematic ( $C_{\text{stat + syst}} = C_\text{stat} + C_\text{syst}$) correlations between supernova measurements.  We see that in Fig.\ref{fig2}, our model-dependent curve shows good agreement with the model-independent PP observation data as well as $\Lambda$CDM model curve across the redshift range.\\

In our analysis, we utilize the Markov Chain Monte Carlo (MCMC) method to constrain cosmological parameters within Narlikar gravity model by analyzing astrophysical observational data, primarily focuse on constrain the free parameter space $(H_0,\Omega_{m0}, \Omega_x, \Omega_x, \Omega_z,w,\xi_0)$ with corresponding the ranges $H_0 \in [60, 80]$, $\Omega_{m0} \in [0.2, 0.7]$, $\Omega_{p} \in [0, 1]$, $\Omega_{q} \in [-0.2, 1]$, $\Omega_{r} \in [0, 1]$, $w \in [-1,1]$ and $\xi_0 \in [0,1]$ respectively. The \texttt{emcee} library \cite{ref36} is employed for parallelized MCMC sampling using 80 walkers and 10000 steps to ensure convergence. By jointly analyzing the  33 observational $H(z)$ data points, and 1048 supernova measurements from the Pantheon compilations, we can extract meaningful constraints on the cosmological parameters and better understand the expansion history of the universe.\\


\section{Results and Disussion}

In this section, we investigate the parameter space ($H_0,\Omega_{m0}, \Omega_x, \Omega_y, \Omega_z,w,\xi_0$) of the Narlikar gravity model with a bulk viscous fluid by jointly analyzing OHD+PP datasets. Our model introduces a creation field, which permits continuous matter generation throughout cosmic evolution. While its impact on the expansion history is subtle, it remains significant: the adopted form of C(t) in (Eq.\ref{eq8}) alters the late-time effective equation of state, producing deviations from $\Lambda$CDM, that can accommodate both local and early-universe values of the Hubble constant. In this contrast, we present the Hubble constant constraint for the Narlikar gravity model with a joint analysis of observational datasets. We obtain a present-day value of  $H_0 = 71.2\pm{2.1} \ \mathrm{km\,s^{-1}\,Mpc^{-1}}$ at 68\% C.L. with  $\Omega_m=0.41^{+0.11}_{-0.09}$ from OHD+PP data sets. This result shows only a mild $0.9\sigma$ discrepancy with SH0ES collaboration’s measurement of $H_0 = 73.27\pm{1.04} \ \mathrm{km\,s^{-1}\,Mpc^{-1}}$  at the 68\% CL., based on supernovae calibrated with Cepheid variables. However, our model successfully alleviates the long-standing $4.1\sigma$ Hubble tension that exists  $5\sigma$ between the Planck collaboration, $H_0=67.36\pm0.54~{\rm km\, s^{-1}\, Mpc^{-1}}$ \cite{ref37}, which is based on CMB observations within the $\Lambda$CDM  model and SH0ES collaboration, $H_0=73.27\pm1.04~{\rm km\, s^{-1}\ Mpc^{-1}}$ \cite{ref37a}, using Cepheid-calibrated Type Ia supernovae.  Hence, our results obtained with Narlikar's gravity model for the Hubble constant, which favour the SH0ES measurement and $\Omega_{m0}$ consistent with Ref\cite{ref37b}.\\

\begin{figure}
    \centering
    \includegraphics[width=1\linewidth]{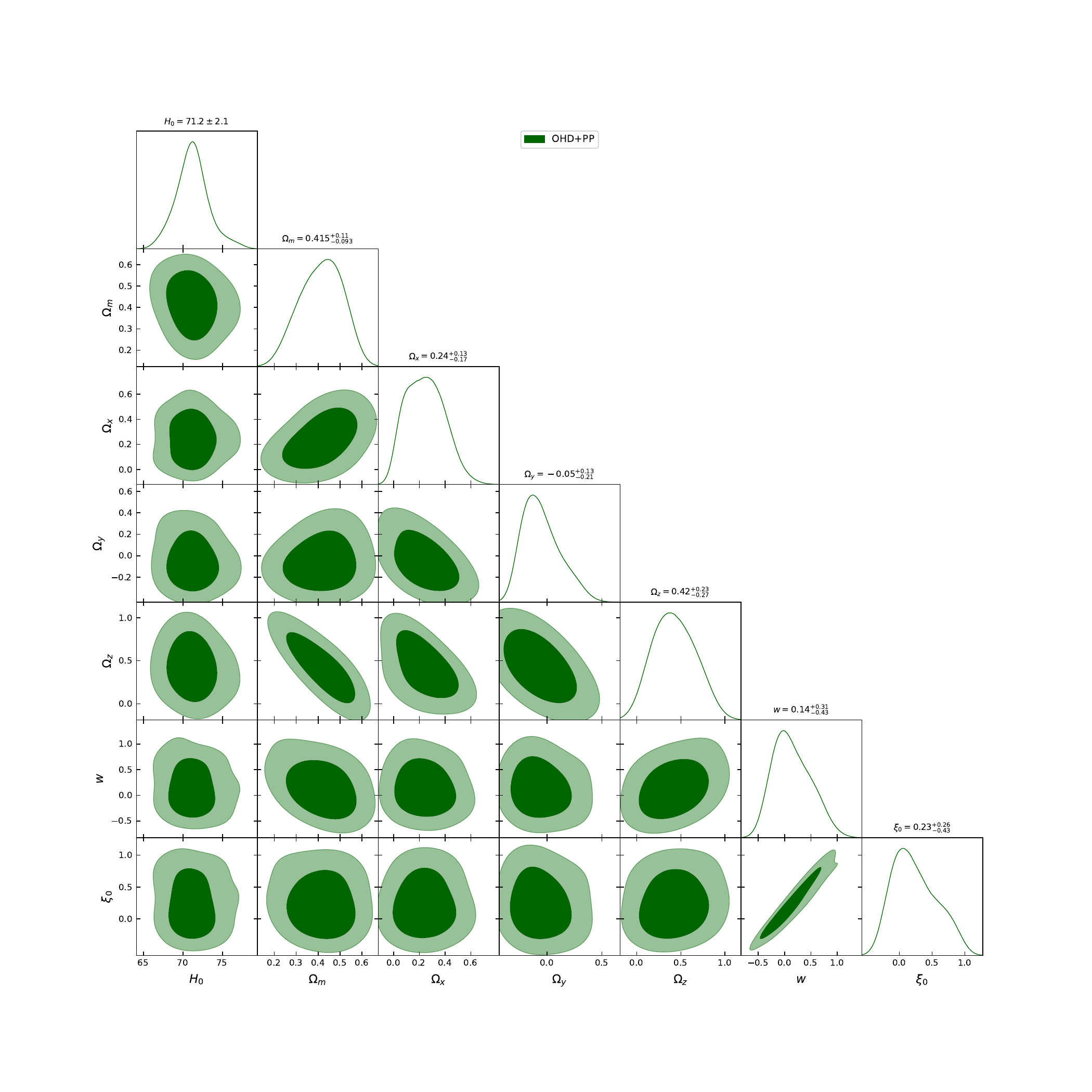}
    \caption{In the triangle plot, we show 1D and 2D contour marginal distribution of free parameter space for the Nalikar gravity model with a bulk viscous fluid from OHD+PP data.}
    \label{fig3}
\end{figure}

From Fig.\ref{fig3}, we notice that the bulk viscosity coefficient $\xi_0$ exhibits a strongly positive correlation with the free parameter $w$, indicating that an increase in $\xi_0$ is associated with an increase in $w$. Also, we find the mean value of the bulk viscosity coefficient as  $\xi_0=0.23$ from the given joint analysis datasets, which remains consistent with values reported in the literature [a,b,c]. In triangle plots, the parameters $H_0$, $\Omega_m$,$\Omega_x$, $\Omega_y$, $\Omega_z$, $w$ and $\xi_0$ are constrained using both 1D marginalized distributions and 2D joint confidence contours, based on the combined datasets OHD+PP.\\

In this scenario, we investigate the sign of the creation field coupling constant ($f$) from various density parameters, $\Omega_x$, $\Omega_y$, and $\Omega_z$.

\begin{itemize}
\item From the analysis of $\Omega_y$ and $\Omega_z$, we established that $\alpha > 0$, implying $\alpha^2 > 0$, and since $H_0^2 > 0$ as well, both terms are strictly positive. Considering  
$\Omega_x = \frac{f \alpha^2}{2H_0^2}\,\frac{(1-w)}{(5+3w-3\xi_0)},$  
and substituting $w = 0.14$ and $\xi_0 = 0.23$, we obtain $\frac{1-w}{5+3w-3\xi_0} \approx 0.1818 > 0$. Hence, the overall multiplicative factor is positive, and the sign of $\Omega_x$ depends solely on $f$. Since the observational value $\Omega_x = 0.24 > 0$, it follows directly that $f > 0$, thereby validating our assumption of a positive $f$ in the analysis.

\item To further check the sign of the coupling constant $f$, we consider the relation $\Omega_y = \tfrac{f \alpha (1+\alpha)}{H_0^2}\tfrac{(w-1)}{4+3w-3\xi_0}$. Since both $\alpha(1+\alpha)$ and $H_0^2$ are strictly positive, their contribution remains positive. For the chosen values $w = 0.14$ and $\xi_0 = 0.23$, the factor $\tfrac{(w-1)}{4+3w-3\xi_0}$ is negative, which makes the overall multiplicative term negative. However, given that $\Omega_y = -0.05 < 0$, the consistency of the relation implies that $f$ must be positive. This result therefore supports and validates our earlier assumption that $f > 0$ in the analysis.

\item To verify the sign of the coupling constant $f$, we use the relation $\Omega_z = \tfrac{f (1+\alpha)^2}{6H_0^2}\tfrac{1-w}{1+w-\xi_0}$. Since both $(1+\alpha)^2$ and $H_0^2$ are strictly positive, their contribution is always positive. For the given values $w = 0.14$ and $\xi_0 = 0.23$, the term $\tfrac{1-w}{1+w-\xi_0}$ also evaluates to a positive quantity, ensuring that the entire multiplicative factor remains positive. As $\Omega_z = 0.42 > 0$, it immediately follows that $f$ must be positive, thereby confirming the validity of our prior assumption that $f > 0$ in the analysis.

\end{itemize}

From the analysis of all three density parameters $\Omega_x$, $\Omega_y$, and $\Omega_z$, it is evident that the creation field coupling constant ($f$) is always greater than zero, thereby confirming the consistency of our assumption $f > 0$ within the Narlikar gravity model with bulk viscous fluid.\\

\textbf{Deceleration Parameter}: The deceleration factor is a key element in cosmology, which displays the behavior of the development of the universe’s expansion rate, which is represented by q(z). The universe's expansion will require a cosmology model that includes its expansion at a slower rate and later, at an increased rate. Hence, it is essential to study the deceleration parameter q, and we write it in terms of H as:

\begin{equation}\label{eq18}
	q(z)=-1+(1+z)\frac{H'}{H}
\end{equation}

\begin{figure}
    \centering
    \includegraphics[width=0.67\linewidth]{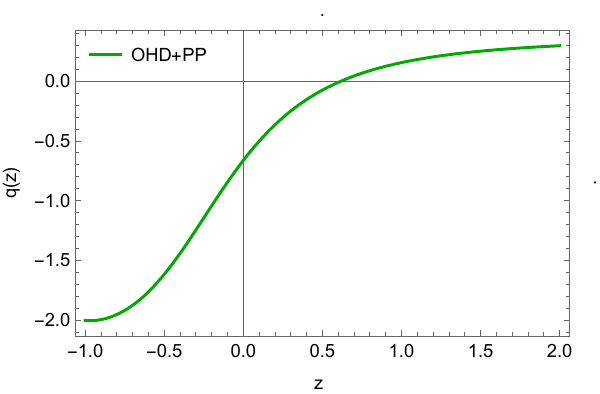}
    \caption{The deceleration parameter $q(z)$ for the combined OHD+PP datasets as a function of redshift z.}
    \label{fig4}
\end{figure}

 The sign of the deceleration parameter (q) indicates that a positive value ($q > 0$) corresponds to a decelerating universe, while the case where q=0 is a critical one, as it denotes a state where the expansion is happening at a constant rate, neither accelerating nor decelerating, which is termed as in equilibrium. In contrast, a negative value of q within the range $-1 < q < 0$ indicates accelerated expansion driven by dark energy–like effects. The special case $q = -1$ corresponds to a pure de Sitter phase, where the expansion of the universe is exponential. If $q < -1$, the expansion becomes super-exponential, signal a “phantom” regime in which the expansion rate grows faster than in de Sitter space.\\

Using the governing equation of  Hubble parameter given in Eq.(\ref{eq12}), and substituting  $H(z)$ and its derivative into Eq.(\ref{eq18}), we obtain the  new expressions for the deceleration parameter 
$q(z)$ presented as Eq.(\ref{eq19}) :

\begin{equation}\label{eq19}
q(z) = -1 + \frac{(1 + z)\left[ 3\Omega_{m0} (1+w-\xi_0)(1 + z)^{3(1+w-\xi_0) - 1} - 2\Omega_x (1 + z)^{-3} - \Omega_y (1 + z)^{-2} \right]}{2 \left[ \Omega_{m0}(1 + z)^{3(1+w-\xi_0)} + \dfrac{\Omega_x}{(1 + z)^2} + \dfrac{\Omega_y}{(1 + z)} + \Omega_z \right]} 
\end{equation}

In this analysis, we present a 2D plot  (Darkgreen curve) of the deceleration parameter $q(z)$ against redshift $z$, which is obtained from the Narlikar model gravity along with the OHD+PP observational data sets shown in Fig.\ref{fig4}. This curve shows increasing trends in the given plot range of redshift. At this stage, the present value of deceleration from OHD+PP data is obtained as $q_0 = 0.65$ for the Narlikar gravity model with bulk viscosity of the fluid. In the same analysis, we find a transition redshift of approximately $z_{tr} \sim 0.63$, which shows the Universe undergoes a fundamental change in cosmic history - the transition from a decelerated expansion phase to the currently observed accelerated expansion of the Universe.  Our obtained result ($q_0$ and $z_{tr}$) are in close agreement with recent results reported in the literature \cite{ref38,ref39,ref40,ref41}.\\

\textbf{Om(z) diagnostic}: The parameter $Om(z)$ diagnostic provided as a powerful null test for all forms of the dark energy (DE) models in this ref. \cite{ref43}. Its direct dependence on the Hubble parameter, $H(z)$, and redshift, $z$, provides an observationally motivated framework for probing the nature of dark energy. This diagnostic provides an efficient way to discriminate between more general dynamical DE models, where the energy density evolves with cosmic time, and the standard cosmological constant model, where the energy density of DE remains fixed. A constant $Om(z)$ across redshifts strongly favors the view that dark energy behaves like the cosmological constant $\Lambda$. On the other side, a redshift-dependent $Om(z)$ diagnostic parameter signals dynamical dark energy behaviour, suggesting a dynamical nature of dark energy, requiring extensions beyond the standard. Specifically, a positive gradient of $Om(z)$ implies phantom dark energy with $ w<-1$, whereas a negative gradient is consistent with quintessence models, for which $-1<w<-\frac{1}{3}$. The mathematical form of the $Om(z)$ diagnostic is

\begin{figure}
    \centering
    \includegraphics[width=0.7\linewidth]{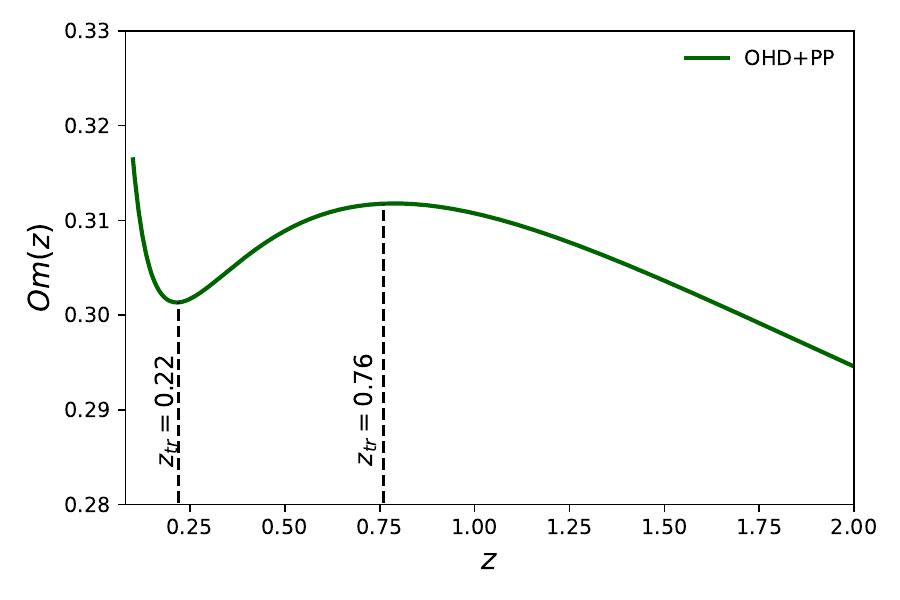}
    \caption{The $Om(z)$ diagnostic for the combined OHD+PP datasets as a function of redshift $z$.}
    \label{fig5}
\end{figure}

\begin{equation}\label{eq20}
Om(z) = \dfrac{\left( \frac{H(z)}{H_0} \right)^2 - 1}{(1+z)^3 - 1}
\end{equation}

from Eq.(\ref{eq12}) and Eq.(\ref{eq20}), we calculate $Om(z)$ in term redshift:
\begin{equation}\label{eq21}
Om(z) = \frac{ \Omega_{m0}(1+z)^{3(1+w-\xi_0)+2} 
+ (\Omega_x+\Omega_y+\Omega_z-1) 
+ z(\Omega_y+2(\Omega_z-1)) 
+ z^2(\Omega_z-1)}
{(1+z)^2 \big[(1+z)^3-1\big]}.
\end{equation}
In this study, we investigates the  $Om$ diagnostic for the Narlikar gravity model with a bulk viscous fluid from the latest combination of observational data (OHD + PP). The behavior of the $Om(z)$ diagnostic, illustrated in Fig.\ref{fig5}, reveals alternating trends with both negative and positive slopes across different redshift regimes. For the $Om(z)$ diagnostic in the low redshift domain, the negative slope is associated with the so-called quiescent type dark energy. In the redshift range $0.22<z<0.76$, the slope is positive, showing a transverse phantom-like behavior. At higher redshifts, the diagnostic reverts to a negative slope, signifying the re-emergence of a quintessence-like phase.\\

In this scenario, our study reveals that only two transition redshift points, which show the transition of the nature of dark energy, are at $z_t = 0.22$, the model predicts a transition from quintessence-phantom foam DE, and at $z_t = 0.76$, a transition from phantom-quintessence behavior occurs. Furthermore, recent studies in the literature indicate that this transition redshift generally lies within the range $0.2 \lesssim z \lesssim 2$ \cite{ref44,ref45}, marking a key epoch in the evolution of dark energy. Our results, obtained using the combined OHD+PP datasets, are in good agreement with this observed range.\\

\textbf{Age of universe}:  We can estimate the age of the universe based on the lookback time, which represents the time interval between the present age of the universe $t_0$ and the age of the universe at a particular redshift
$z$. Since redshift reflects the degree of cosmic expansion, it naturally serves as a variable to trace the universe’s past, and its relation with the Hubble parameter H(z) provides a rigorous framework for evaluating cosmic timescales. In a cosmological model where the Hubble parameter $H(z)$ varies with redshift, the lookback time is mathematically expressed as
\begin{equation}
t_0 - t(z) = \int_0^z \frac{dz'}{(1 + z') H(z')}
\label{eq22}
\end{equation}

We obtain the present age of the universe from   Eq.(\ref{eq22}) by employing the Narlikar cosmological model given in Eq.(\ref{eq12}). In this formulation, the beginning of the Universe is characterized by $t=0$, which corresponds to the limit \( z \to \infty \). Consequently, the present cosmic age is determined by evaluating the integral over the entire redshift range, we get

$$ H_0 t_0 = \int_0^{\infty} \frac{dz'}{(1+z') \sqrt{\Omega_{m0}(1 + z)^{3(1+w -\xi_0)} + \frac{\Omega_{p}}{ (1 + z)^2} +  \frac{\Omega_{q}}{ (1 + z)} +\Omega_r}}$$

 \begin{figure}[hbt!]
     \centering
     \includegraphics[width=0.7\linewidth]{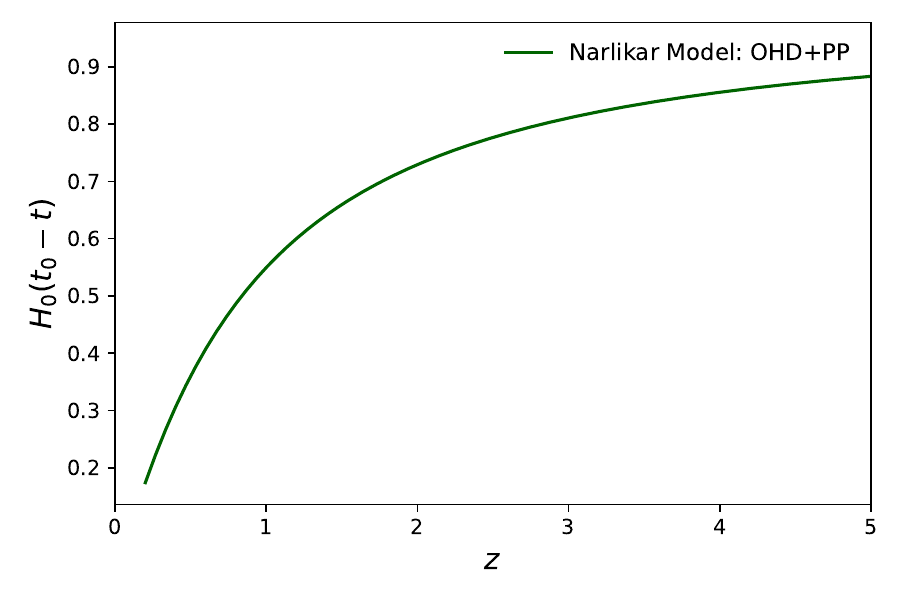}
     \caption{The age of the universe for the combined OHD+PP datasets as a function of redshift $z$.}
     \label{fig6}
 \end{figure}

In this analysis, we present a two-dimensional plot of the normalized lookback time, $H_0(t_0-t)$ as a function of redshift z in
Fig.\ref{fig6}, derived from our Narlikar gravity model using the combined OHD+PP observational datasets. The dark green
curve of lookback time indicates a clear increasing trend across the largest redshift interval.
For the parameter values $H_0 = 71.2 \pm 2.1 \ \text{km s}^{-1} \text{Mpc}^{-1}$,
$\Omega_{m} = 0.41^{+0.11}_{-0.09}$, $\Omega_{x} = 0.24^{+0.13}_{-0.17}$, $\Omega_{y} = -0.05^{+0.13}_{-0.21}$, $\Omega_{z} = 0.42^{+0.23}_{-0.27}$, $w = 0.14^{+0.31}_{-0.43}$ and $\xi_0 = 0.23^{+0.26}_{-0.43}$, the corresponding present age of the universe is calculated by   \textbf{Narlikar gravity model:} {$t_{0} \approx 13.50 \pm 1.80  \mathrm{Gyr}$}. This result is in good agreement with recent independent determinations reported in the literature, such as $13.6\pm 0.2$ Gyrs \cite{ref45}, $13.50\pm 0.23$ Gyrs \cite{ref46}.\\

\textbf{Energy Condition}: In cosmology,$T_{\mu}{\nu}$, which represents the distribution and flow of energy and momentum in spacetime. These conditions are fundamental in general relativity and play a crucial role in studying the evolution and structure of the universe. They are often used to derive general theorems, such as the singularity theorems by Hawking and Penrose.\\

The next step is to verify whether the obtained solution derived from considering Narlikar's gravity model remains physically reasonable by verifying the energy conditions. These conditions provide rules that the energy–momentum tensor must obey to keep the energy density positive. Some relevant energy conditions—NEC, DEC, and SEC—are derived from the Raychaudhuri equation and take the form [67] and mathematics expressed as:

\begin{itemize}

\item Null Energy Condition (NEC) =   $\rho+p\geq0$
\item Dominant Energy Condition (DEC) = $\rho-p\geq0$
\item  Strong Energy Condition (SEC) = $\rho+3p\leq0$

\end{itemize}
By applying the above energy conditions, we can examine the feasibility of our models. Furthermore, this analysis provides deeper insights into the realistic behavior of our universe. The corresponding set of energy conditions, expressed in terms of 
density parameter and $z$, are obtained as

\begin{equation}\label{eq23}
\rho + p = H_0^2 \left[ 
3(1+w)\Omega_{m0}(1+z)^{3(1+w-\xi_0)} 
+ (3\xi_0 - 2)\frac{\Omega_x}{(1+z)^2} 
+ (3\xi_0 - 1)\frac{\Omega_y}{(1+z)} 
+ 3\xi_0 \Omega_z 
\right] 
+ f\left(1 + \frac{\alpha z}{1+z}\right)^2
\end{equation}

\begin{equation}\label{eq24}
\rho - p = H_0^2 \left[
3(1-w)\,\Omega_{m0}(1+z)^{3(1+w-\xi_0)}
+ (8 - 3\xi_0)\,\frac{\Omega_x}{(1+z)^2}
+ (7 - 3\xi_0)\,\frac{\Omega_y}{(1+z)}
+ 3(2 - \xi_0)\,\Omega_z
\right]
\end{equation}

\begin{equation}\label{eq25}
\rho + 3p = H_0^2 \left[
3(1+3w)\,\Omega_{m0}(1+z)^{3(1+w-\xi_0)}
+ (9\xi_0 - 12)\,\frac{\Omega_x}{(1+z)^2}
+ (9\xi_0 - 9)\,\frac{\Omega_y}{(1+z)}
+ (9\xi_0 - 6)\,\Omega_z
\right]
+ 2f\left(1 + \frac{\alpha z}{1+z}\right)^2
\end{equation}

\begin{figure}[hbt!]
    \centering
    (a)\includegraphics[width=0.47\linewidth]{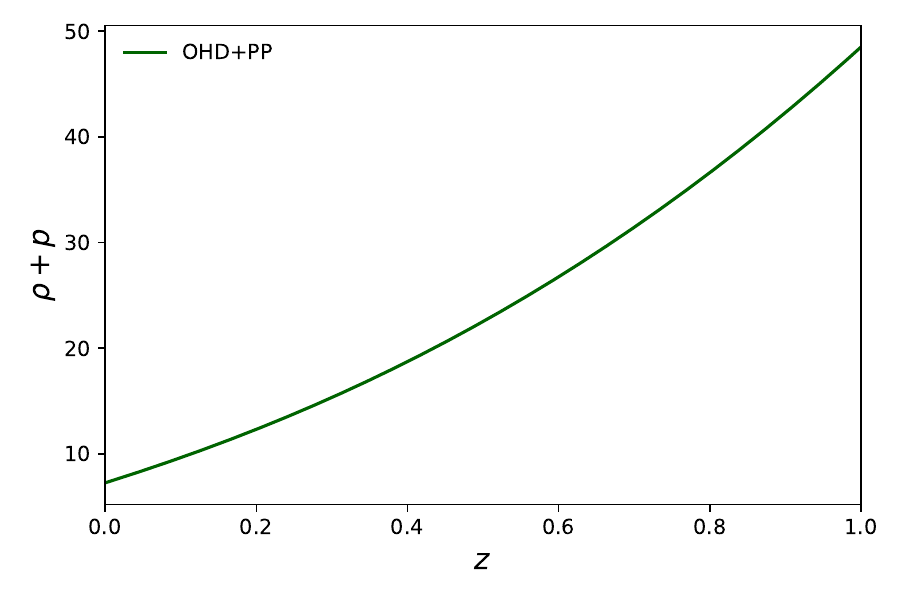}
   (b) \includegraphics[width=0.47\linewidth]{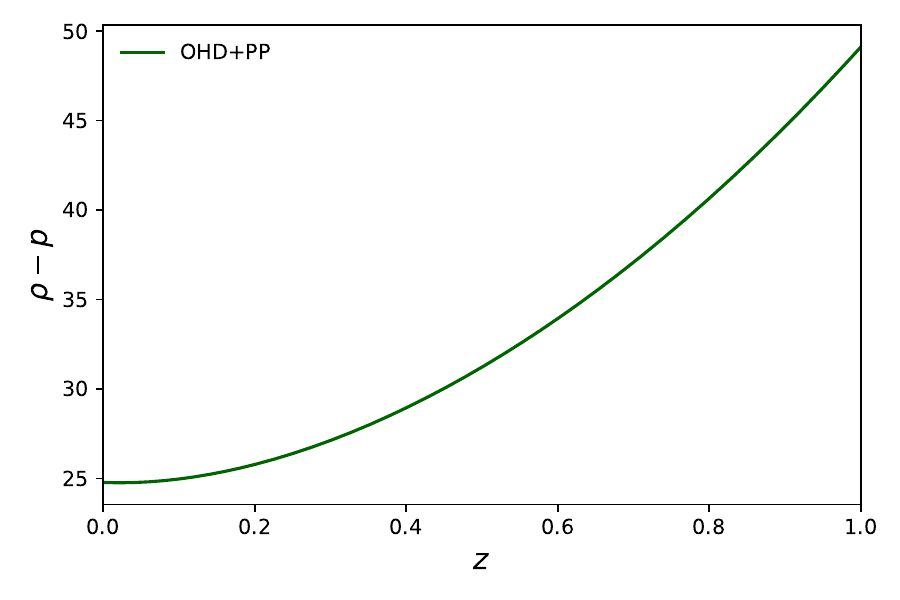}
   (c) \includegraphics[width=0.48\linewidth]{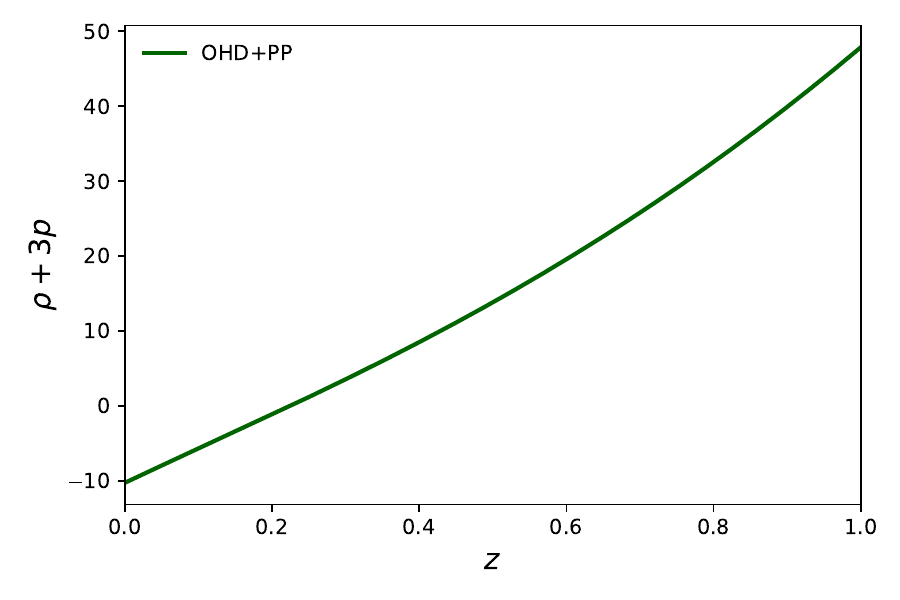}
\caption{Normalized plots of the energy conditions as functions of $z$. 
    (a): NEC ($\rho+p$), (b): DEC ($\rho-p$), (c): SEC ($\rho+3p$).}
    \label{fig7}
\end{figure}

Fig.\ref{fig7} (a \& b) highlight that the NEC and DEC are satisfied for the full range of redshifts. Hence, the model maintains physical admissibility throughout its cosmic evolution. However,  Fig.\ref{fig7}(c) indicates a violation of the SEC in the present era, consistent with the universe’s accelerated expansion.

\section{Conclusion}
In this work, we have presented a comprehensive investigation of the Hoyle–Narlikar (HN) gravity
model in the presence of a bulk viscous cosmic fluid, employing a newly proposed functional form of 
the creation field $C(t) = t + \int \alpha (1 - a)\,dt + c_1$.Within this framework, we derived the 
modified Friedmann equations and the continuity equation incorporating both the viscous pressure 
term and the dynamical creation field contribution. By solving these equations analytically, we 
obtained the general behavior of the energy density  $\rho(a)$, the Hubble parameter $H(z)$, and the
effective thermodynamic quantities $\rho_{eff}$ and $p_{eff}$This formulation ensures that the model
remains analytically tractable while allowing for dynamical flexibility, thereby offering a robust 
theoretical setup to explore the late-time cosmic dynamics in the context of the HN theory.\\

A key achievement of this study lies in the joint observational analysis using the latest 33-point Observational Hubble Data (OHD) and the Pantheon Supernovae (PP) compilation. The best-fit values obtained from the combined OHD+PP datasets indicate $H_{0}=71.2\pm2.1$ $MpC^{-1}$ and $\xi_{0}=0.23$
and a positive creation field coupling constant  $f$. The model predicts a transition redshift of $\sim z_{t}0.63$ , marking the onset of the present cosmic acceleration phase, with the estimated age of the Universe $13.50\pm1.80$ Gyr. The positive and statistically constrained value of $f$ confirms the active role of the creation field in driving cosmic dynamics, while the inclusion of bulk viscosity contributes to effective negative pressure, assisting the acceleration mechanism. Notably, the slightly higher  $H_{0}$ value inferred from this analysis suggests a potential alleviation of the well-known  $\sim 4.1\sigma$  Hubble tension between local and early-Universe measurements.\\

Furthermore, the analysis of the energy conditions shows that the model remains stable and consistent with the requirement for accelerated expansion at late times. The combination of a bulk viscous fluid and a dynamical creation field acts effectively as a unified dark energy component, naturally reproducing the observed cosmic acceleration without introducing additional exotic fields or modifications to the standard relativistic framework. This highlights the physical viability of the model in explaining the late-time cosmic acceleration and the evolving nature of dark energy within a geometrically modified framework.\\

Overall, our results demonstrate that the Narlikar gravity model with bulk viscous fluid and the new creation field form can successfully accommodate observational constraints while providing a coherent theoretical alternative to the standard $\Lambda$CDM scenario. This framework offers a promising avenue to probe unresolved cosmological puzzles—such as the Hubble tension, the origin of dark energy, and the mechanism of cosmic acceleration—within the broader context of creation-field cosmology. Future work can extend this study by exploring perturbation dynamics, structure growth, and cosmic microwave background constraints within this model, which would further clarify its consistency with the full suite of cosmological observations and its role as a potential alternative to standard cosmology.

\begin{acknowledgments}
\noindent 
The authors (A. Pradhan \& A. Dixit) are thankful to IUCAA, Pune,
India for providing support and facility under Visiting
Associateship program.  M. Yadav is supported by a
Junior Research Fellowship (CSIR/UGC Ref. No. 180010603050) from the University Grants Commission, Govt. of India


\end{acknowledgments}

\end{document}